\documentclass[preprint]{vgtc}               

\graphicspath{{figures/}{pictures/}{images/}{./}} 

\usepackage{times}                     
\usepackage{booktabs}
\usepackage{csquotes}
\usepackage{mathptmx}
\usepackage{enumitem}

\onlineid{6718}

\vgtccategory{Research}

\vgtcinsertpkg

\preprinttext{%
  \centering
  \parbox{0.95\textwidth}{%
    \centering
    \textcopyright\ 2024 IEEE. Presented at the 2024 IEEE International Symposium on Mixed and Augmented Reality Adjunct \\
    (ISMAR-Adjunct). The published article is available at: \href{https://ieeexplore.ieee.org/abstract/document/10765177}{https://ieeexplore.ieee.org/abstract/document/10765177} DOI:\href{https://doi.org/10.1109/ISMAR-Adjunct64951.2024.00166}{10.1109/ISMAR-Adjunct64951.2024.00166}%
  }%
}

\begin{document}

\title{VR Cloud Gaming UX: Exploring the Impact of Network Quality on Emotion, Presence, Game Experience and Cybersickness}





\author{Maximilian Warsinke\thanks{e-mail: warsinke@tu-berlin.de}\\ %
    \parbox{1.6in}{\scriptsize \centering Quality and Usability Lab, Technische Universität Berlin, Berlin, Germany}
\and Tanja Kojić\thanks{e-mail: tanja.kojic@tu-berlin.de}\\ %
     \parbox{1.6in}{\scriptsize \centering Quality and Usability Lab, Technische Universität Berlin, Berlin, Germany}
\and Maurizio Vergari\thanks{e-mail: maurizio.vergari@tu-berlin.de}\\ %
     \parbox{1.6in}{\scriptsize \centering Quality and Usability Lab, Technische Universität Berlin, Berlin, Germany}
\and Jan-Niklas Voigt-Antons\thanks{e-mail: jan-niklas.voigt-antons@hshl.de}\\ %
    \parbox{1.6in}{\scriptsize \centering Immersive Reality Lab, Hochschule Hamm-Lippstadt, Lippstadt, Germany}
\and Sebastian Möller\thanks{e-mail: sebastian.moeller@tu-berlin.de}\\ %
    \parbox{1.6in}{\scriptsize \centering Quality and Usability Lab, Technische Universität Berlin, \\ DFKI, Berlin, Germany}}

\abstract{
This study explores the User Experience (UX) of Virtual Reality (VR) cloud gaming under simulated network degradation conditions. Two contrasting games (Beat Saber, Cubism) were streamed via Meta Air Link to a Quest 3 device in a laboratory setup. Packet loss and delay were introduced into the streaming network using NetEm for Wi-Fi traffic manipulation. In a within-subjects experiment, 16 participants played both games under three network conditions (Loss, Delay, Baseline), followed by post-game questionnaires assessing their emotions, perceived quality, game experience, sense of presence, and cybersickness. Friedman's test and Dunn's post-hoc test for pairwise comparisons revealed that packet loss had a greater impact on UX than delay across almost all evaluated aspects. Notably, packet loss in Beat Saber led to a significant increase in cybersickness, whereas in Cubism, players experienced a significant reduction in their sense of presence. Additionally, both games exhibited statistically significant variations between conditions in most game experience dimensions, perceived quality, and emotional responses. This study highlights the critical role of network stability in VR cloud gaming, particularly in minimizing packet loss. The different dynamics between the games suggest the possibility of genre-specific optimization and novel game design considerations for VR cloud games.
} 



\maketitle
\section{Introduction} \label{introduction}
Recent VR headsets have been predominately designed as standalone devices that enable users to run applications directly on an integrated CPU. Nonetheless, resource-intensive simulations and high-end games still require a personal computer (PCVR). In VR cloud gaming, the game runs on a cloud server and is streamed directly to the device. Although great possibilities emerge with externalized computing power, the bottleneck shifts towards network connection. This may lead to new challenges for ensuring the optimal UX.

While cloud gaming for mobile games is already well-researched, few studies on VR cloud gaming exist. First steps to assess the Quality of Experience (QoE) in the area were done by Li et al. by simulating a cloud connection through a local network \cite{li_performance_2020}. Results indicate that especially low bandwidth and a high packet loss rate harmed the QoE. Delay had a lower effect, but affected games that required high responsiveness (Beat Saber). These results provide an opportunity to further investigate UX dynamics. Zhao et al. used cloud computing with a server hosted outside the lab to create a traffic dataset \cite{zhao_virtual_2021}. The results identify latency and data rate limit as crucial factors for QoE.

The study presented in this paper aims to explore the UX in VR cloud gaming by employing full-length questionnaires on a comprehensive selection of metrics. All the aspects investigated were selected under the expectation of being influenced by network degradation.
One psychological metric to approach the UX is emotion. As modelled with the circumplex model of affect, players' emotions can be quantified using the two dimensions of valence and arousal \cite{Russell_1980}.
According to Slater and Wilbur, presence is defined as the subjective quality experienced by feeling present in a virtual environment \cite{slater1997framework}. Although wearing a head-mounted display (HMD) enhances this feeling, enjoyment is not an immediate consequence \cite{lemmens_fear_2022}. Shelstad et al. found that players preferred the VR version when comparing the same game to their computer monitor version \cite{shelstad_gaming_2017}. However, this result could not be replicated by other researchers with different types of games \cite{yildirim_video_2018, carroll_effects_2019}, highlighting the importance of genre variability.
Measuring game experience is a complex endeavour with many impacting factors. Schaffer and Fang conducted an extensive literature review on game enjoyment \cite{Schaffer_Fang_2019}, discussing aspects like challenge, immersion, engagement, competition, self-determination, and motivation. One pivotal aspect is the difficulty level of the game, which should correspond to the player's abilities to induce a flow state \cite{csikszentmihalyi_flow_1991}. This state is often described as the optimal game experience desired by game designers.
In VR environments, users may experience a form of motion sickness known as cybersickness with symptoms like headaches, nausea, and eyestrain \cite{LaViola_2000}. Sensory mismatch between visual and vestibular inputs is the most widely accepted theory for its origin \cite{Gallagher_Ferrè_2018}. Yildirim et al. found that game experience is negatively impacted by cybersickness \cite{Yildirim_2019}.
Combining these findings with VR cloud gaming revealed a research gap, leading to the following research question:

\begin{itemize}[itemsep=0pt,topsep=0pt]
    \item [RQ:] To what extent do packet loss and delay affect users' UX aspects (quality, emotion, presence, game experience, cybersickness) when playing two different VR games streamed via Wi-Fi?
\end{itemize}

\begin{figure}[ht]
\centering
\includegraphics[width=0.23\textwidth]{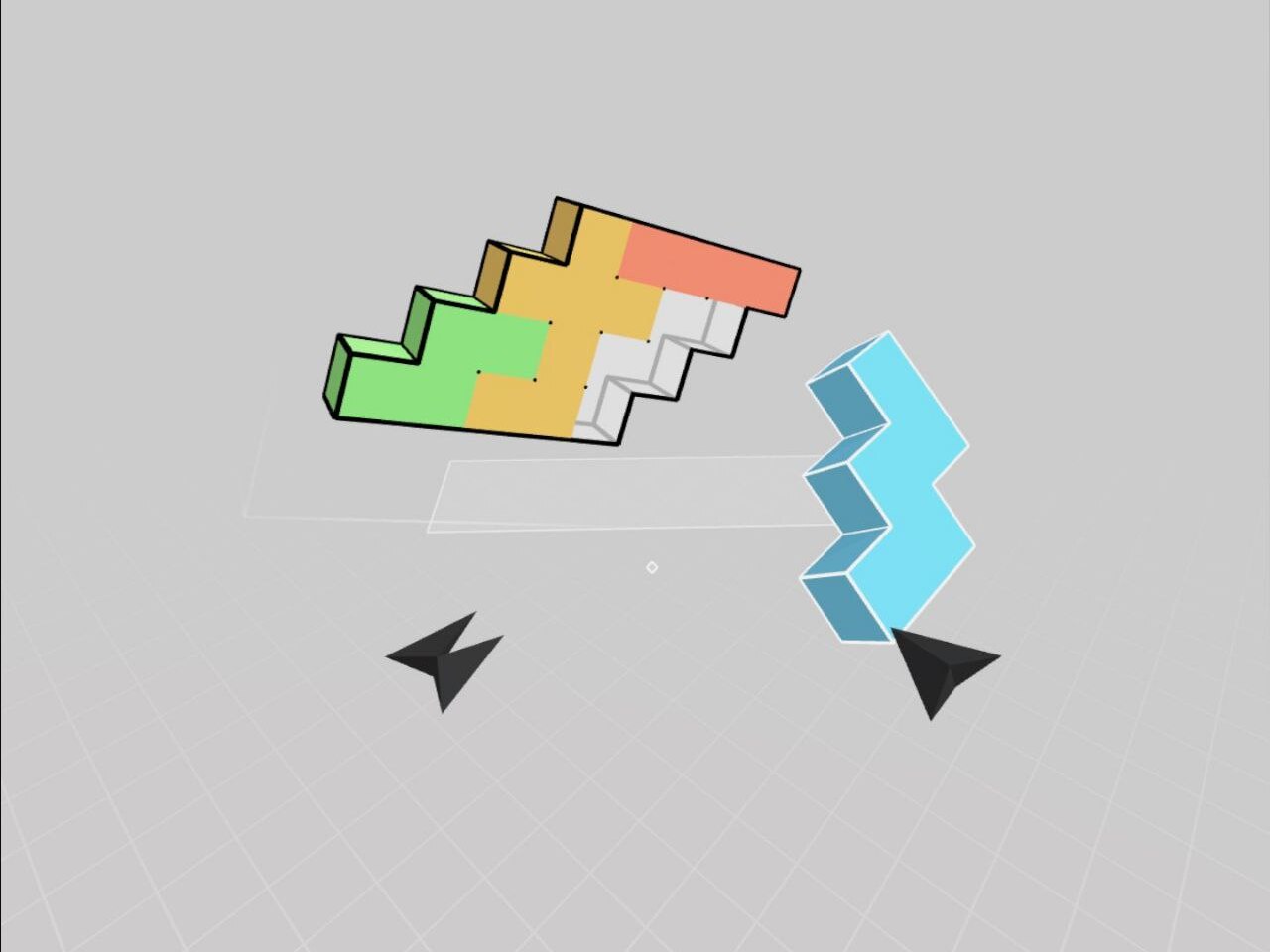} 
\includegraphics[width=0.23\textwidth]{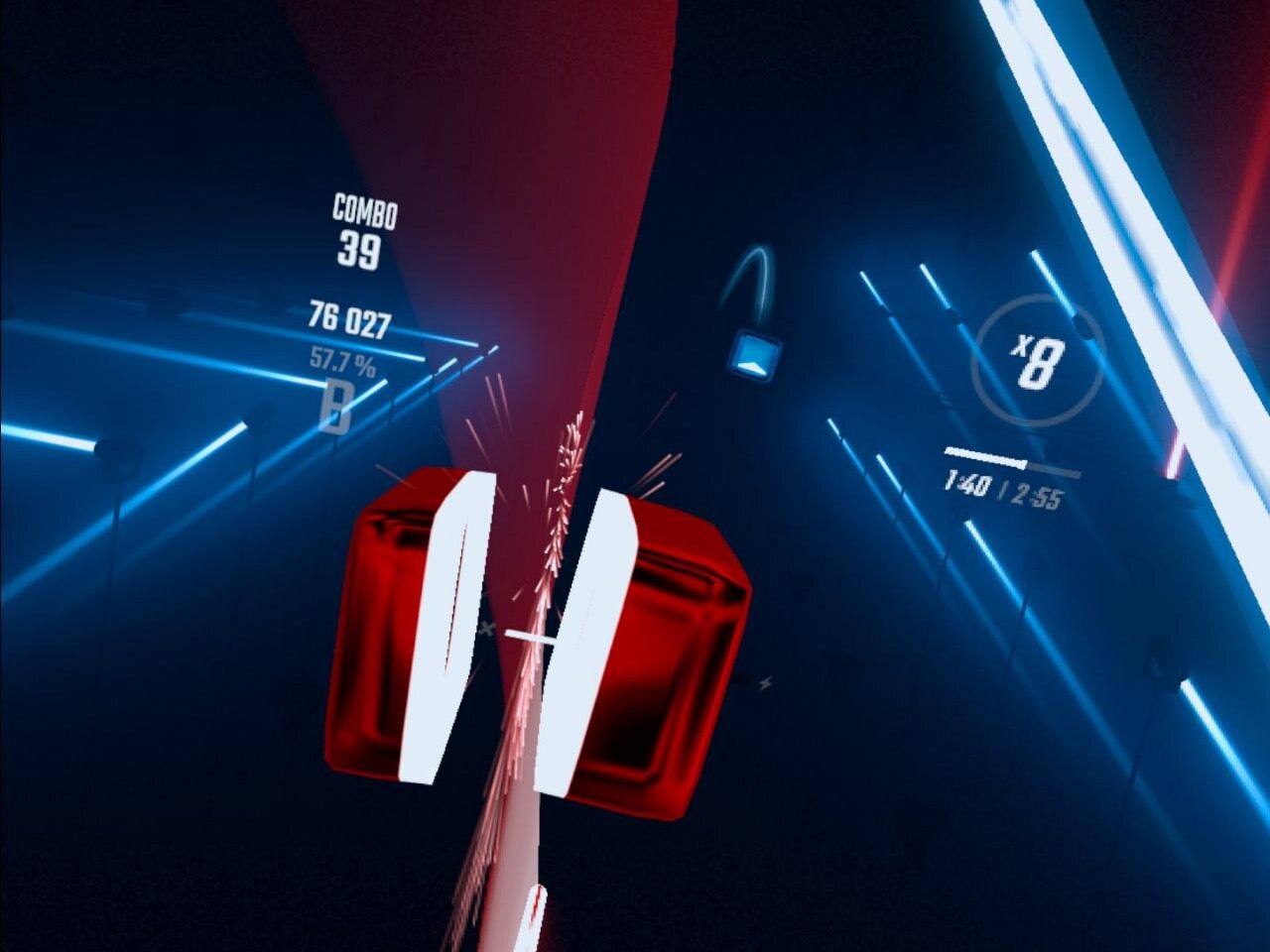}\\
\vspace{-1em}
\caption{Cubism (left) and Beat Saber (right).}
\label{fig:fig_games}
\vspace{-1em}
\end{figure}

In this user study, participants were invited to play two VR games streamed over a local network under the influence of packet loss and delay. Post-condition questionnaires were employed to assess perceived quality, emotion, presence, cybersickness, and game experience. To gain a more holistic impression, two games with contrasting gameplay were chosen. Cubism, a spatial puzzle game, and Beat Saber, a fast-paced rhythm game (Figure \ref{fig:fig_games}).

\section{Methods}
\subsection{Participants}
A total of 16 people participated in the study. Among them were 7 females and 9 were male. The average age was 29.5 (SD=3.72), and 9 of them indicated that there were students. Regarding eyesight improvement, 6 participants wore glasses and 2 lenses. For the experiment, 1 person removed their glasses. The self-reported experience with VR was 2.25 on average (SD=1.18), whereas the average experience with gaming was 3.12 (SD=1.31). The experience of VR gaming was 1.81 on average (SD=0.83) (all scales ranged from 1 to 5, with 5 being the highest). Among the participants, 5 already played the game Beat Saber, and 1 played Cubism. The Affinity for Technology Interaction (ATI) score was 4.43 (SD=0.99).

\subsection{Experimental Design}
A within-factor design was used for the experiment. Each participant played through 6 conditions (A1, A2, A3, B1, B2, B3) in 3-minute game sessions, which were randomized in advance to prevent sequence effects. After each condition, the participants were asked to complete post-condition questionnaires. The experiment was approved by the local ethics committee of Technical University of Berlin.

\vspace{2mm}
\noindent
Game:
\begin{itemize}[itemsep=-1pt,topsep=2pt]
    \item A: Beat Saber
    \item B: Cubism
\end{itemize}
Quality Setting:
\begin{itemize}[itemsep=-1pt,topsep=2pt]
    \item 1: Baseline
          (0\% loss, 0ms delay)
    \item 2: Delay
          (0\% loss, 40ms delay)
    \item 3: Loss
          (0.3\% loss, 0ms delay)
\end{itemize}
\vspace{2mm}

\subsection{Technical Setup}
A desktop computer featuring an NVIDIA GTX1080, Intel(R) Core(TM) i7-7700K CPU @ 4.20GHz, and 16.0GB RAM was used to simulate a cloud connection. Using an Apple AirPort Extreme 802.11ac router, together with a Fujitsu Lifebook S935, a local network was created for the experiment, which could be manipulated to simulate different network conditions using NetEM \cite{hemminger2005network}. The PC was tethered to the router with an Ethernet cable, whereas the HMD was connected via Wi-Fi (5GHz) at a distance of 2 m. As HMD, the Meta Quest 3 was chosen. It is equipped with a Snapdragon XR2 Gen 2 GPU and 8GB DRAM and can display a resolution of 2064 × 2208 pixels per eye. For streaming, Meta Air Link (v63.0) was used with an update rate of 72 Hz, as recommended by the system, and a render resolution of 4128 × 4416 pixels. A fixed bitrate of 80Mbit/s was chosen.

\subsection{Games}
The decision to use Beat Saber for testing was made for comparison with other studies in the field (cf. Section \ref{introduction}). Cubism was chosen to complement Beat Saber. In both games, players find themselves in an abstract environment with geometric shapes as game objects. Additionally, both games are renounced from a narrative, strengthening the focus in comparing the gameplay.

\noindent
\textbf{Beat Saber}:
Beat Saber\footnote{\url{https://beatsaber.com/}} is a VR rhythm game that encourages full-body movement gameplay. A sequence of geometric objects corresponding to the rhythm of an audio track is generated. The player must perform slashing actions with virtual lightsabers to destroy the objects. For the experimental sessions, a song with a duration of 3 minutes was selected (\enquote{Curtains (All Night Long)}) that all participants played in the easiest difficulty setting.\\
\noindent
\textbf{Cubism}:
Cubism\footnote{\url{https://www.cubism-vr.com/}} is a VR puzzle game that challenges the player with spatial 3D riddles. The objective is to sort geometrical shapes into a predefined form by arranging the floating objects directly with the controllers. For the experimental sessions, the level progress was reset ahead of every condition. The participants played for 3 minutes from the beginning.

\subsection{Conditions}
A pre-test was conducted with two colleagues to test the conditions and determine the values for network traffic control. The following conditions were selected for the experiment.

\noindent
\textbf{Baseline}: The Baseline condition is an optimal streaming experience that functions as the reference point.

\noindent
\textbf{Delay}: A fixed delay of 40 ms was added using NetEm. The Delay condition resulted in a minimal but constant delay. Delays can disrupt the real-time feedback and potentially lead to failure within the game.

\noindent
\textbf{Loss}: A packet loss rate of 0.3\% was introduced using NetEm. The Loss condition results in artificial lags and stutters that periodically disturb the game. Packet loss can hinder the responsiveness and fluidity of the game.

\subsection{Questionnaires}
\begin{itemize}[itemsep=-1pt,topsep=3pt]
    \item Demographics: age, gender, profession, vision impairments, previous experience with VR, and previous experience with gaming (ranging from 1 to 5).
    \item Affinity for Technology Interaction (ATI) \cite{franke2019personal} (ranging from 1 to 6).
    \item Overall Quality: Mean Opinion Score (MOS) (ranging from 1 to 7).
    \item Emotion: Self Assessment Mannequin (SAM)\cite{bradley1994measuring} (\textit{Valence}, \textit{Arousal}, \textit{Dominance} ranging from 1 to 5).
    \item Presence: iGroup Presence Questionnaire (IPQ) \cite{igroup} (\textit{Spatial Presence}, \textit{Involvement}, \textit{Realness}, \textit{Overall Presence} ranging from 1 to 7).
    \item Cybersickness: Cybersickness in Virtual Reality Questionnaire (CSQ-VR) \cite{CSQ-VR} (\textit{Nausea, Vestibular, Oculomtor} ranging from 2 to 14, \textit{Total} ranging from 6 to 32).
    \item Game Experience: In-game Game Experience Questionnaire (In-game GEQ) \cite{geq} (\textit{Compentence, Immersion, Tension, Flow, Challenge, Positive Affect, Negative Affect} ranging from 0 to 4, item 1 was changed to item 12 from the Core GEQ following the ITU-T recommendation P.809 \cite{itu2018}).
    \item Final questionnaire: preferred game and general feedback.
\end{itemize}

\subsection{Procedure}
An introduction to the experiment was followed by two short training sessions (one per game). After the training, the participants were asked to complete the initial questionnaires (Demographics, ATI, and CSQ-VR). The experiment started with one of the six conditions (randomized). The participants played under the quality condition for 3 minutes and were then asked to fill out the post-condition questionnaires (MOS, SAM, GEQ, IPQ, and CSQ-VR). This procedure was repeated for all the six conditions. After the last condition, a final questionnaire was added, and the participants received 15€ compensation.

\subsection{Statistical Testing}
Due to violations of normality and a relatively small sample size, Friedman's tests were conducted for each post-condition questionnaire's scores instead of repeated measures ANOVA. If a statistically significant difference was found ($p<0.05$), Dunn's post-hoc test for pairwise comparisons was used. The $p$-values were then adjusted using Bonferroni correction for multiple comparisons.

\section{Results}

\subsection{Overall Quality}
\begin{figure}[ht]
\hspace*{-1em}
\centering
\includegraphics[width=0.45\textwidth]{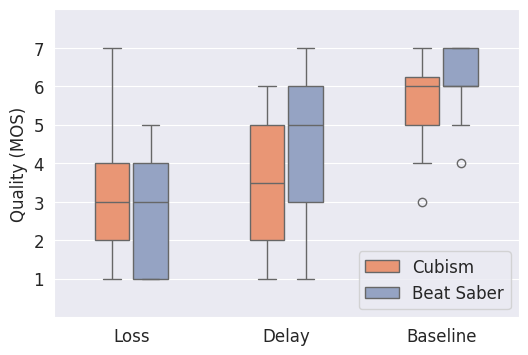}
\caption{MOS scores grouped by game and condition (Whiskers denote distribution outside quartiles, circles indicate outliers).}
\label{fig:fig_mos}
\vspace{-1em}
\end{figure}

MOS scores were statistically significantly different for both games. Post-hoc tests revealed significant decreases for Beat Saber under the Loss condition (M=2.62) compared to Baseline (M=6.12). In Cubism, both the Delay (M=3.38) and the Loss condition (M=3.38) were significantly lower compared to the Baseline (M=5.69) (all $p$-values $<$0.001). The quality scores for each game and condition are visualized in Figure \ref{fig:fig_mos}.

\subsection{Emotion}
\textit{Valence} and \textit{Dominance} differed significantly for both games. Post-hoc, the Loss condition (Beat Saber M=2.94, Cubism M=3.25) showed a significant decrease in \textit{Valence} compared to the Baseline condition (Beat Saber M=4.62, Cubism M=4.31) (Beat Saber $p$=0.001, Cubism $p$=0.040). For \textit{Dominance} only a statistically significant decrease between Loss (M=2.88) and Baseline (M=4.12) for Beat Saber ($p$=0.006) was found.

\subsection{Presence}
In Cubism, significant differences were found for \textit{Spatial Presence}, \textit{Involvement}, and \textit{Overall Presence}. In Beat Saber, only \textit{Overall Presence} showed a significant difference. Post-hoc, \textit{Spatial Presence} in Cubism was significantly lowered for both Delay (M=4.99, $p$=0.040) and Loss (M=4.71, $p$=0.001) compared to the Baseline condition (M=5.66). \textit{Involvement} was significantly lower for Cubism in Loss (M=4.17) compared to the Baseline condition (M=5.48) ($p$=0.006).

\subsection{Game Experience}
\textit{Competence}, \textit{Immersion}, \textit{Flow}, \textit{Tension}, \textit{Challenge} and \textit{Positive Affect} were significantly different for both games. \textit{Negative Affect} was only significantly different for Beat Saber. Post-hoc, for Beat Saber, \textit{Competence} was significantly lower in Loss (M=1.67) than in the Baseline condition (M=3.10) ($p$=0.001). For Cubism, \textit{Competence} was significantly lower in Delay (M=2.00) than in Baseline (M=3.00) ($p$=0.024). \textit{Immersion} was significantly lower for both games in Loss (Beat Saber M=2.27, Cubism M=2.03) compared to the Baseline condition (Beat Saber M=3.13, Cubism M=3.00) (Beat Saber $p$=0.041, Cubism $p$=0.003). \textit{Flow} was significantly lower in Cubism between Loss (M=2.00) and Baseline (M=2.97) ($p$=0.006). In both games \textit{Tension} was significantly higher in Loss (Beat Saber M=2.47, Cubism M=2.40) compared to Baseline (Beat Saber M=0.43, Cubism M=0.67) (Beat Saber $p$=0.001, Cubism $p$=0.002). \textit{Positive Affect} was significantly lower for Beat Saber in Loss (M=1.83) compared to the Baseline condition (M=3.37) ($p<$0.001). For Cubism, \textit{Positive Affect} was significantly lower for both Loss (M=1.90, $p<$0.001) and Delay (M=2.33, $p$=0.041) compared to Baseline (M=3.33).

\subsection{Cybersickness} 
\begin{figure}[ht]
\centering
\includegraphics[width=0.45\textwidth]{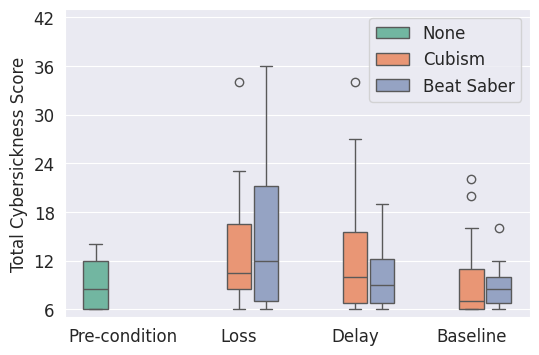}
\caption{Total scores from CSQ-VR grouped by game and condition (Whiskers denote distribution outside quartiles, circles indicate outliers).}
\label{fig:fig_csq}
\end{figure}

Significant differences were found for Beat Saber in the \textit{Nausea}, \textit{Vestibular}, and \textit{Total} scores, while for Cubism \textit{Oculomotor} scores showed significant differences. Post-hoc, \textit{Nausea} in Beat Saber was significantly higher in Loss (M=4.96) compared to the Baseline condition (M=2.62) ($p$=0.024). The \textit{Total} score for Beat Saber is significantly higher in Loss (M=14.62) than in the Pre-Condition (M=8.94) and Baseline condition (M=8.75) (both $p$=0.045). The \textit{Total} cybersickness scores are visualized for each game and condition in Figure \ref{fig:fig_csq}.

\section{Discussion \& Conclusion}
This study explored the influence of packet loss and delay on the UX of streamed VR games with contrasting gameplay. 

It was found that delay only impacted Cubism, contrary to expectations (cf. Section \ref{introduction}). It is assumed that players of Beat Saber adapted to an offset of 40 ms since the delay was fixed. In Cubism, the delay significantly affected \textit{Positive Affect}, \textit{Competence}, and \textit{Spatial Presence} negatively. This allows the assumption that the enjoyment of playing is reduced, as indicated by the scores for \textit{Positive Affect} and \textit{Competence}. The reduced \textit{Spatial Presence} could be derived from the movement offset, which also lowers the precision when arranging the building blocks. Thus, it may be less consistent with the physical world experience.

Packet loss was perceived in both games, as indicated by the MOS scores, and negatively affected many of the tested UX dimensions for both games. The player's \textit{Valence} was more negative compared to the baseline, and in Beat Saber, participants felt less in control, as shown through \textit{Dominance}. In terms of game experience, packet loss had an impact on \textit{Competence}, \textit{Immersion}, \textit{Tension}, and \textit{Positive Affect} for both games and on \textit{Flow} only for Cubism. This indicates that the game experience is heavily disturbed and that the stutters caused by packet loss create frustration, as shown by \textit{Tension}. \textit{Spatial Presence} and \textit{Involvement} are significantly reduced only for Cubism under packet loss, which supports the assumption that quality degradations in Cubism easily disrupt the plausibility of the virtual world. Cybersickness was only significantly increased for Beat Saber in Loss, specifically \textit{Nausea}, which also affected the \textit{Total} cybersickness score. This is likely because Beat Saber is in constant movement that cannot be stopped when stutters from packet loss occur. In Cubism, participants often waited for the quality to stabilize before continuing to play. This aligns with the findings of Pöhlmann et al. that more visual motion causes more cybersickness \cite{Pöhlmann_2023}. They also found that cognitive load could reduce cybersickness, which could indicate that Cubsim might require more mental effort.

It was shown that network quality plays a crucial role in the UX when streaming VR games. Although VR cloud gaming has many practical advantages, it does not occur without its pitfalls. In addition to the reduced enjoyment of the game experience, quality degradation can induce discomfort in players. Cybersickness resulting from stutter due to network instability can be a dealbreaker for streaming VR games, and the feeling of presence, a pivotal reason for VR usage, can be reduced by delays. The findings highlight opportunities for optimization in VR cloud gaming and invite genre-specific considerations in performance prioritization and design.

\section{Future Work}
Since the participants played the same games repeatedly, learning effects may have influenced the game experience. These effects were mitigated by randomizing the conditions. For cybersickness scores, carryover effects from previous conditions cannot be ruled out, because symptoms can last up to multiple hours \cite{Stanney_Kennedy_1998}.
In the future, more distinct network settings and more gradual packet loss and delay values could provide more insights. Testing with a variable delay instead of a fixed delay could yield promising results that may be closer to real-world conditions. A more in-depth analysis of cybersickness values could provide further insights, such as a subsequent analysis over time and consideration of participants' demographics, as age and gender are often discussed as influential factors for symptom strength \cite{Dilanchian_Andringa_Boot_2021, Poehlmann_Li_McGill_Pollick_Brewster_2023}.

\bibliographystyle{abbrv-doi}

\bibliography{main}
\end{document}